\documentclass[12pt]{article}

\usepackage{psfig}

\title{\huge XAS Study of the High Pressure\\ Behaviour of Quartzlike
Compounds}

\author{James {\sc Badro}, Jean-Paul {\sc Itié}, Alain {\sc Polian} and
Philippe {\sc Gillet}}

\date{August 1, 1996}

\begin{document}
\maketitle

{\bf \centerline{Abstract}}

\begin{center}
\begin{minipage}{14cm}

\footnotesize
\renewcommand{\baselinestretch}{0.8}

EXAFS spectroscopy experiments have been carried out on quartz-like
$\alpha -$GaAsO$_4$ and $\alpha -$AlAsO$_4$ at high pressure and room
temperature. It
has been shown that these materials exhibit two structural phase
transitions; the first transition to a high pressure crystalline form
occurs at 9 GPa and is reversible upon decompression, whereas the
second transition occurs at higher pressures and is irreversible. In
$\alpha -$GaAsO$_4$, EXAFS measurements agree with the predicted transition
from four- to six-fold coordination of oxygen atoms around the cations,
but the two local coordination transformations are not dissociated; in
fact, both As and Ga atoms exhibit a coordination change at the onset
of the first phase transition, the rate of transformation being
significantly higher for Ga atoms.
In both cases, the average bond length increases very rapidly with
pressure thus yielding the first compression stage after the
transition. In the second stage, the average bond lengths increase
slowly, ultimately reaching six-fold coordination above 28 GPa and 24
GPa for As and Ga respectively. The behaviour of the As K-edge EXAFS is
the same for both compounds, and enables us to link the behaviour of Ga
and Al atoms. The local transformations are well described and a direct
link with phosphate berlinites seems timely.

\end{minipage}
\end{center}

\vspace{4mm}
{\bf \Large Introduction}
\vspace{4mm}

Quartz and quartz-like compounds are among nature's most interesting
minerals, because their various structures in different regions of
thermodynamic phase space are simple yet common ones. The low pressure
polymorphs consist of tetrahedrally coordinated cations whereas the
higher pressures forms have octahedral cation arrangements, in which
oxygen atoms are packed in a body-centred cubic (bcc) sublattice around
the cations. At room temperature, the kinetics of these four- to
six-coordinated transitions are very slow leading to a pressure-induced
amorphization (PIA) process in quartz. This phenomenon has been
observed by different probes \cite{hemley2,kingma1,hazen1,williams2}
on the SiO2 system.
The room temperature compression and PIA of isostructural $\alpha
-$GeO$_2$ has also been studied by X-ray diffraction and EXAFS
spectroscopy \cite{itie1} and the study has shown that germanium
coordination increases from four to six through the
crystal-to-amorphous transition. Molecular dynamics (MD)
simulations \cite{tse1,chelikowsky2,bing2,watson1,badro1} have shown
 that changes in the local structure during
the transition in quartz are very similar to those observed in
GeO$_2$.
In quartz-like berlinites, it was first thought that $\alpha -$AlPO$_4$
exhibited the same room temperature and high pressure behaviour as
quartz \cite{kruger1}, and underwent PIA at 15 GPa. A later Raman
scattering study \cite{gillet2}
as well as Brillouin spectroscopy measurements \cite{polian1}
have shown that this
transition is not a PIA, but a polymorphic crystalline phase
transition, the high pressure form being very poorly crystallized, and
therefore yielding very poor X-ray diffraction patterns. MD simulations
associate the transition with a destabilization of the AlO$_4$ tetrahedron
and its transition towards an AlO$_6$ octahedron;
they also have concluded that PIA would occur only when the PO$_4$
tetrahedra became unstable, at a predicted pressure of 80 GPa, which
has not yet been produced experimentally using AlPO$_4$. In-situ EXAFS
experiments were performed \cite{itie2} on an isostructural compound,
GaPO$_4$, at
the Ga K-edge, and the polymorphic phase transition observed at 12 GPa
is associated with a four to six-fold oxygen coordination change around
gallium atoms.

\vspace{4mm}
{\bf \large Experimental}
\vspace{4mm}

Quartz-structured aluminium and gallium arsenate powders were used for
this high pressure dispersive EXAFS study. The samples were loaded in a
stainless steel inconel gasket with a 250 mm hole diameter and 50 mm
thickness and high pressures were generated by a membrane driven
diamond anvil cell (DAC). Silicon oil was used as a pressure
transmitting medium, and ruby fluorescence was used as a pressure
gauge. The measurements were carried out on the D-11
dispersive \cite{dartyges1,tolentino1}
EXAFS beamline on the DCI storage ring of the LURE synchrotron facility
in Orsay, France. Each absorption measurement was accumulated for 32
times and 3.5 seconds each, and was followed by the collection of a
spectrum of a GaAs sample in order to insure edge energy stability.

\vspace{4mm}
{\boldmath \large $GaAsO_{4}$}
\vspace{4mm}

EXAFS spectra at the Ga and As K-edges show a very interesting pattern
with increasing pressure (fig. 1); a first polymorphic transition \cite{itie3}
occurs around 9 GPa and is observed for both Ga and As. The average
Ga--O distance rises rapidely and then levels off above 12 GPa
corresponding to an octahedral arrangement of gallium atoms. Meanwhile,
the As--O mean distance also  increases between 9 and 12 GPa at which
point another compression regime appears and the As--O bond lengthens in
a slower manner, until a plateau is reached at about 25 GPa. These
distance variations are related to the rate of transformation from
4-fold to 6-fold coordinated cations; the arsenic atoms have two
transformation regimes, the second being much slower because the
six-coordinated structure of gallium atoms renders the system less
compressible.
This would therefore imply that the coordination of both As and Ga
atoms transform at 9 GPa, and that the Ga 4-to-6 transformation is
faster, permitting the As 4-to-6 transformation to continue up to 25
GPa. At this pressure, the system consists of a pseudo-rutile type
structure, the XANES spectra for As and Ga atoms being very similar to
the XANES spectrum of GeO$_2$ at high pressure \cite{itie1}. A recent X-ray
diffraction study \cite{clark1} by Clark et al. has shown that the intermediate
crystalline phase can be indexed in a triclinic (yet very close to
orthorhombic) structure. It was pointed out that this phase has 2/3 of
the cations in 6-fold coordination and the rest in 4-fold. Our EXAFS
measurements concur with this result and it can be added that in this
particular case, all Ga atoms are in six-fold coordination whereas a
smaller portion of As  atoms are in octahedral coordination.

\vspace{2mm}
\begin{figure}[h]
\begin{minipage}{4cm}
\centerline{\psfig{figure=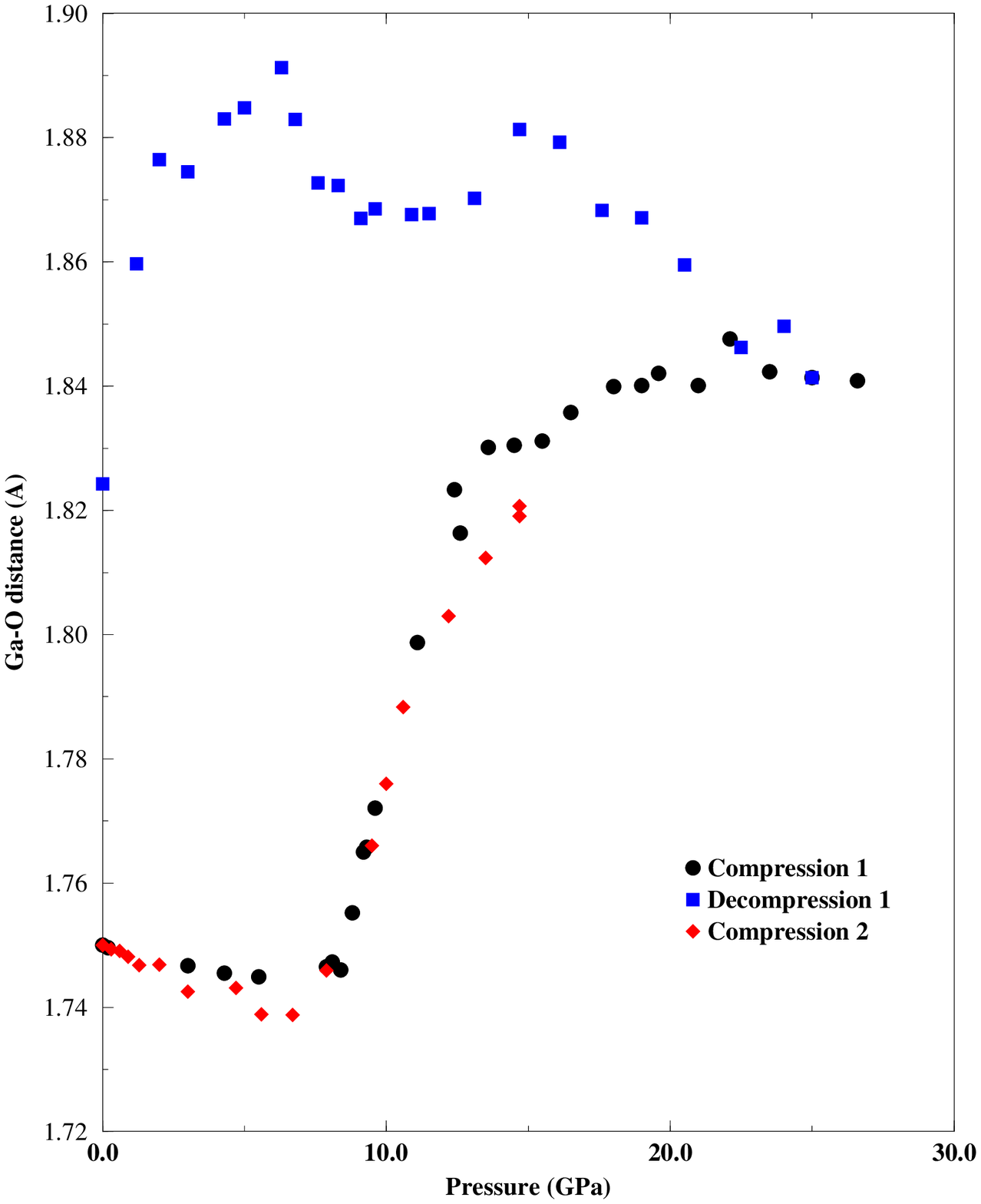,height=6cm,width=3.5cm}}
\caption{Ga--O mean distance as a function of pressure obtained
by EXAFS in $GaAsO_{4}$}
\label{gaga}
\end{minipage}
\hfill
\begin{minipage}{4cm}
\centerline{\psfig{figure=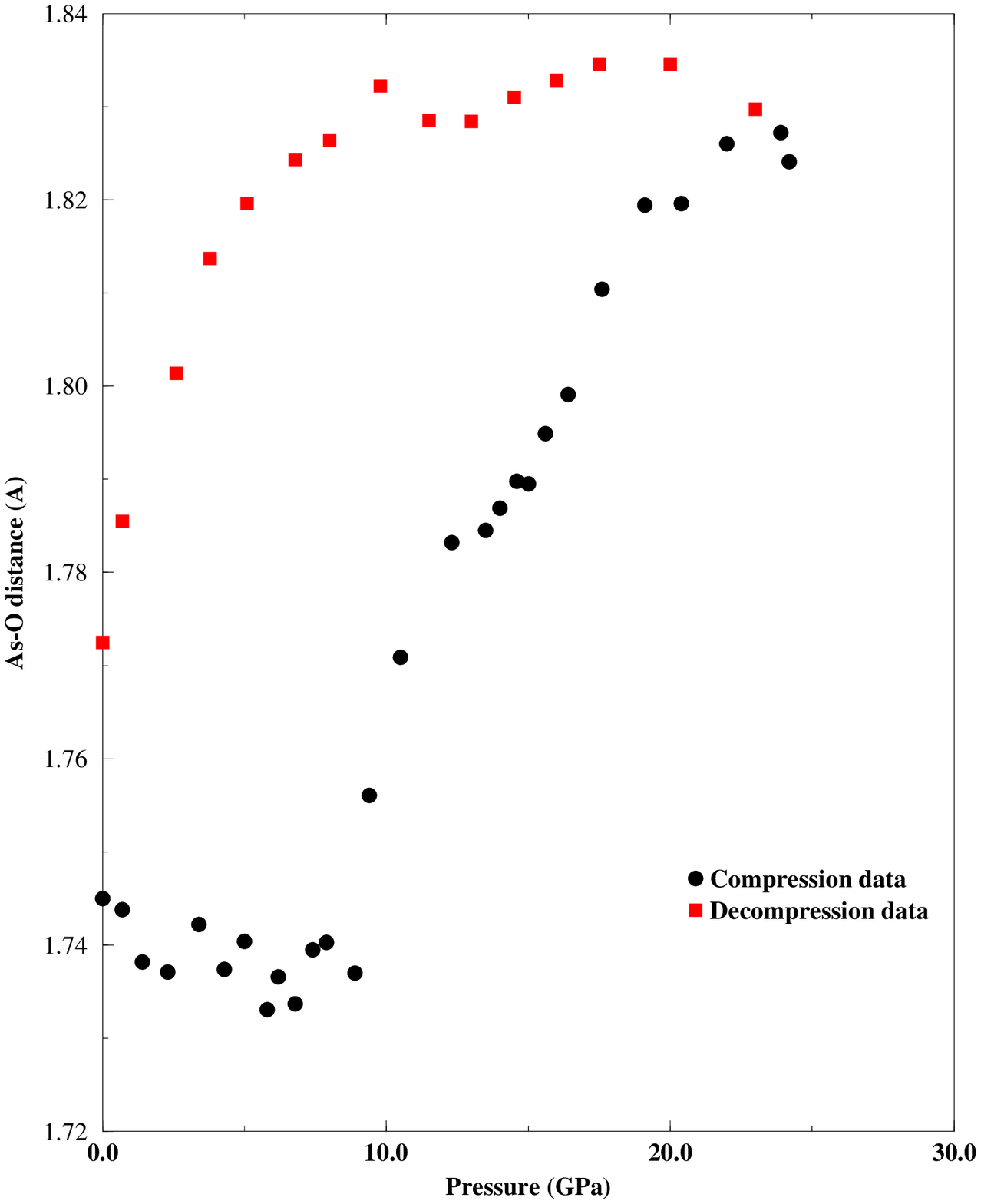,height=6cm,width=3.5cm}}
\caption{As--O mean distance as a function of pressure obtained
by EXAFS in $GaAsO_{4}$}
\label{gaas}
\end{minipage}
\hfill
\begin{minipage}{4cm}
\centerline{\psfig{figure=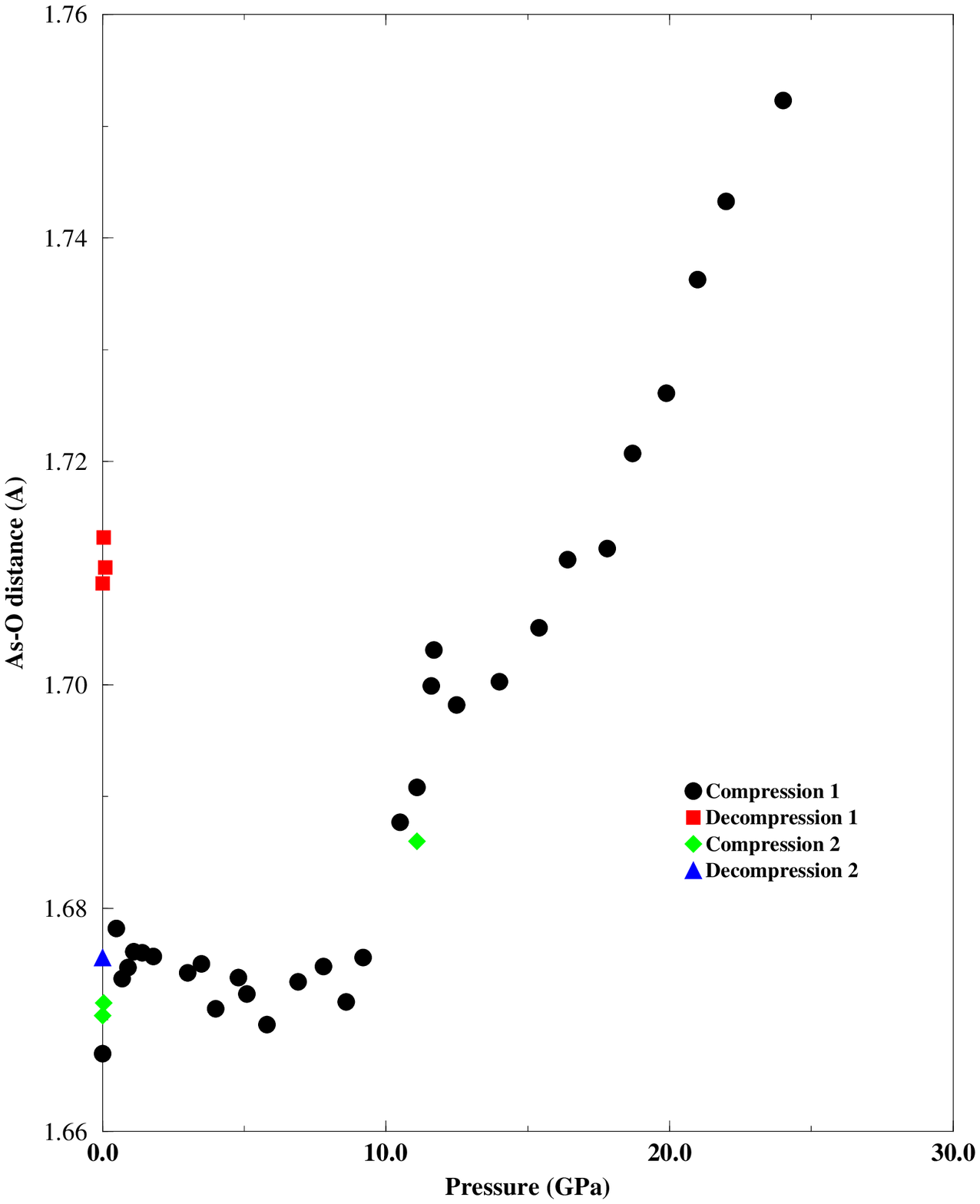,height=6cm,width=3.5cm}}
\caption{As--O mean distance as a function of pressure obtained
by EXAFS in $AlAsO_{4}$}
\label{alas}
\end{minipage}
\end{figure}

In a GaAsO$_4$ sample quenched from 15 GPa, after all Ga atoms have
become six-fold coordinated and before all As clusters turn into
octahedra, the Ga atoms undergo a back-transformation to a tetrahedral
arrangement with partial amorphization of the sample, whereas the
sample decompressed from 25 GPa has a much more complex history and
shows amorphous character after decompression, mainly consisting of
octahedral configurations.

\vspace{4mm}
{\boldmath \large $AlAsO_{4}$}
\vspace{4mm}

A recent X-ray diffraction study \cite{sowa3} has shown
that AlAsO$_4$ undergoes a
crystal-to-crystal phase transition between 7.6 and 8.4 GPa. This study
also claimed that the high pressure crystalline phase undergoes PIA
between 8.4 and 9.4 GPa. Our X-ray diffraction and Raman scattering
experiments disagree with these conclusions. In fact, Raman scattering
on single crystals as well as X-ray powder diffraction clearly agree
with the first crystal-to-crystal transition between 8.5 and 9.5 GPa,
but crystalline reflections of the high pressure phase are observed up
to 35 GPa and Raman spectra show new crystalline features between 9.6
and 13.3 GPa.
The As K-edge EXAFS show a rapid increase in As coordination between 9
and 12 GPa (fig. 2) followed by a lightly less rapid rise in As--O
distance up to 28 GPa, exactly as seen in GaAsO$_4$.  A direct
comparison with GaAsO$_4$ in which both cations can be monitored by
in-situ EXAFS, would imply that Al coordination should rise rapidly
from 4 to 6 and should level off at about 12 GPa at which point only As
coordination increases ultimately reaching a value of 6. Upon
decompression, the transition is reversible below 13 GPa and becomes
irreversible above this pressure.

\vspace{4mm}
{\bf \large Conclusion}
\vspace{4mm}

We have shown that both aluminium and gallium arsenate undergo two
phase transitions at high pressure and 300 K. The first transformation
is a crystalline polymorphic phase transition and is associated, as
opposed to phosphates \cite{itie3}, with a local
transition from four- to six-fold
coordination of both cations. The second transition occurs over a large
pressure domain, for it is not a usual first order transformation, but
a PIA, and starts slightly above the point where one of the cations
reaches complete six-fold coordination. As with quartz, samples
decompressed from these pressure domains are more or less amorphous,
depending on the maximum pressure reached. The first transition is
locally reversible yet producing some amorphous material, whereas the
second transition leading to a 'pressure glass' is not.
The destabilization of tetrahedral clusters in both GaAsO$_4$ and
AlAsO$_4$ has been demonstrated by in-situ EXAFS, and is in good
agreement with MD studies on berlinite. Experimental and numerical
studies are converging on the reversibility of the first transition;
the remaining tetrahedral clusters (PO$_4$ for phosphates, and AsO$_4$ for
arsenates) seem to form an undeformed sublattice that forces the
transformed octahedra back into their original positions when pressure
is released. In the case of arsenates, if these strong clusters are
transformed into octahedra, the back-transformation is not possible for
the driving force has vanished, and the systems becomes amorphous. In
the case of phosphates, the critical pressure needed to destabilize the
PO$_4$ tetrahedra has not yet been reached.

\newpage
\footnotesize
\renewcommand{\baselinestretch}{0.8}
\bibliographystyle{unsrt}

\begin{thebibliography}{10}

\bibitem{hemley2}
R.J. Hemley, A.P. Jephcoat, H.K. Mao, L.C. Ming, and M.H. Manghnani.
\newblock {\em Nature}, {\bf 334}:52--54, 1988.

\bibitem{kingma1}
K.J. Kingma, C.~Meade, R.J. Hemley, H.~Mao, and D.R. Veblen.
\newblock {\em Science}, {\bf 259}:666--669, 1993.

\bibitem{hazen1}
R.M. Hazen, L.W. Finger, R.J. Hemley, and H.K. Mao.
\newblock {\em Solid state communications}, {\bf 72}(5):507--511, 1989.

\bibitem{williams2}
Q.~Williams, R.J. Hemley, M.B. Kruger, and R.~Jeanloz.
\newblock {\em J. Geophys. Research}, {\bf 98}(B12):22,157--22,170, 1993.

\bibitem{itie1}
J.-P. Itié, A.~Polian, G.~Calas, J.~Petiau, A.~Fontaine, and H.~Tolentino.
\newblock {\em Phys. Rev. Lett.}, {\bf 63}(9):389--401, 1989.

\bibitem{tse1}
J.S. Tse and D.D. Klug.
\newblock {\em Phys. Rev. Lett.}, {\bf 67}(25):3559, 1991.

\bibitem{chelikowsky2}
J.R. Chelikowsky, H.E. King, N.~Troullier, J.L. Martins, and J.~Glinnemann.
\newblock {\em Phys. Rev. Let.}, {\bf 65}:3309--3312, 1990.

\bibitem{bing2}
N.~Binggeli and J.R. Chelikowsky.
\newblock {\em Nature}, {\bf 353}:344--346, 1991.

\bibitem{watson1}
G.W. Watson and S.C. Parker.
\newblock {\em Philosophical Mag. Lett.}, {\bf 71}(1):59--64, 1995.

\bibitem{badro1}
J.~Badro, J.-L. Barrat, and Ph. Gillet.
\newblock {\em Phys. Rev. Lett.}, {\bf 76}(5):772--775, 1996.

\bibitem{kruger1}
M.B. Kruger and R.~Jeanloz.
\newblock {\em Science}, {\bf 249}(647):647--649, 1990.

\bibitem{gillet2}
Ph. Gillet, J.~Badro, B.~Varrel, and P.F. McMillan.
\newblock {\em Phys. Rev. B}, {\bf 51}(17):11262--11269, 1995.

\bibitem{polian1}
A.~Polian, M.~Grimsditch, and E.~Philippot.
\newblock {\em Phys. Rev. Lett.}, {\bf 71}(19), 1993.

\bibitem{itie2}
J.-P. Itié, A.~Polian, D.~Martinez, V.~Briois, A.~DiCicco, A.Filipponi, and
  A.~San Miguel.
\newblock {\em J. Physique (Paris)}, {\bf }(Accepted), 1997.

\bibitem{dartyges1}
E.~Dartyges, C.~Depautex, J.M. Dubuisson, A.~Fontaine, A.~Jucha, P.~Leboucher,
  and G.~Tourillon.
\newblock {\em Nucl. Inst. Meth.}, {\bf A246}:456, 1986.

\bibitem{tolentino1}
H.~Tolentino, E.~Dartyges, A.~Fontaine, and G.~Tourillon.
\newblock {\em J. Appl. Phys.}, {\bf 21}:15, 1988.

\bibitem{itie3}
J.-P. Itié, T.~Tinoco, A.~Polian, G.~Demazeau, S.~Matar, and E.~Philippot.
\newblock {\em High Pressure Research}, {\bf 14}:269--276, 1996.

\bibitem{clark1}
S.M. Clark, A.G. Christy, R.~Jones, J.~Chen, J.M. Thomas, and G.N. Greaves.
\newblock {\em Phys. Rev. B}, {\bf 51}(1):38--51, 1995.

\bibitem{sowa3}
H.~Sowa and H.~Ahsbahs.
\newblock {\em Zeit. f\"ur Krist.}, {\bf }(211):96--100, 1996.

\end{thebibliography}

\normalsize

\end{document}